\documentclass[runningheads]{llncs}

\usepackage[utf8]{inputenc} 
\usepackage[T1]{fontenc}    
\usepackage{hyperref}       
\usepackage{url}            
\usepackage{booktabs}       
\usepackage{amsfonts}       
\usepackage{nicefrac}       
\usepackage{microtype}      
\usepackage{lipsum}
\usepackage{graphicx}

\usepackage{newunicodechar}
\newunicodechar{ﬁ}{fi}
\newunicodechar{ﬀ}{ff}
\begin{document}
\title{A Medical Literature Search System for Identifying Effective Treatments in Precision Medicine}
%
%
\author{Jiaming Qu\and
Yue Wang}
\authorrunning{J. Qu and Y. Wang}
%
\institute{School of Information and Library Science \\ University of North Carolina at Chapel Hill \\ Chapel Hill, NC \\
\email{jiaming@ad.unc.edu, wangyue@email.unc.edu}}
\maketitle              
\begin{abstract}
The Precision Medicine Initiative states that treatments for a patient should take into account not only the patient’s disease, but his/her specific genetic variation as well. The vast biomedical literature holds the potential for physicians to identify effective treatment options for a cancer patient. However, the complexity and ambiguity of medical terms can result in vocabulary mismatch between the physician’s query and the literature. The physician’s search intent (finding treatments instead of other types of studies) is difficult to explicitly formulate in a query. Therefore, simple ad hoc retrieval approach will suffer from low recall and precision.

In this paper, we propose a new retrieval system that helps physicians identify effective treatments in precision medicine. Given a cancer patient with a specific disease, genetic variation, and demographic information, the system aims to identify biomedical publications that report effective treatments. We approach this goal from two directions. First, we expand the original disease and gene terms using biomedical knowledge bases to improve recall of the initial retrieval. We then improve precision by promoting treatment-related publications to the top using a machine learning reranker trained on 2017 Text Retrieval Conference Precision Medicine (PM) track corpus. Batch evaluation results on 2018 PM track corpus show that the proposed approach effectively improves both recall and precision, achieving performance comparable to the top entries on the leaderboard of 2018 PM track. 
\end{abstract}
\section{Introduction}
Mr. H is a 35-year-old man who has had a cough for a couple of months. After having an x-ray and CT scan, he was diagnosed as having non–small cell lung cancer. However, he has always kept a healthy lifestyle and never smoked. One possible answer could be that his disease is caused by some gene mutations. Had his cancer been diagnosed in the past, his physician might have treated him just like other patients with lung cancers caused by smoking. But if his disease is diagnosed recently, his biopsy tissue could be analyzed for a panel of genetic variants that can reliably predict what treatment would be most effective for him \cite{mirnezami2012preparing}.

This scenario reveals the core idea of \emph{precision medicine}: based on genetic, biomarker and demographic data, finding a personalized treatment which is different from treatments for other patients with similar clinical symptoms for a specific patient. According to the Precision Medicine Initiative launched by the former President Barack Obama in 2015, precision medicine is an approach for disease treatment and prevention that takes into account individual variability in genes, environment, and lifestyle for each person. This approach will allow physicians to conduct treatments more accurately. It is in contrast to the one-size-fits-all approach, in which treatments are developed for all the people with the same disease, with less consideration for the differences between individuals \cite{collins2015new}. One example of precision medicine is finding treatments according to the genes of that patient, as genomic sequencing can be used as a molecular microscope to classify disease according to their specific abnormal biology. A study supports that in 96\% of undiagnosed primary tumors, a genomic alteration could be identified and that in 85\% of these cases, it is potentially treatable by a known drug \cite{ashley2015precision}.

The origin of precision medicine could date back to the fact that a person who needs a blood transfusion is not given blood from a randomly selected donor, but it was impossible to fulfill this goal in the past. The reduced cost of sequencing a human genome (\$22 million ten years ago versus \$1,000 to \$5,000 now) and time of sequencing (2 years ten years ago versus 1 day now) has made the precision medicine idea more possible today \cite{hudson2015precision}. Also, the adoption of Electronic Health Record (EHR) systems in hospitals has grown from below 30\% to over 90\%. These systems can help physicians to find relevant and useful information about a patient to inspire and support their decisions. Such information is helpful in evidence-based analysis. In early days, physicians would input patient’s demographic data and symptoms, searching for information in a EHR system to conduct disease diagnosis. They analyze the data to determine the pathophysiologic explanation, or to find potential treatment by referring to previous treatments for patients with similar symptoms \cite{musen2014clinical}.

Although such kind of analysis which is based on previous evidence and records may still work today, in most cases however, academic literature could provide more comprehensive knowledge and scientific support for decision-making with more reliability, and physicians can learn about the latest research of and treatments for a disease as well. The ability to personalize treatment in a scientifically rigorous manner is thus the hallmark of the emerging precision medicine paradigm. However, due to the huge volume of academic articles which keeps growing rapidly, it is not easy to build an effective Information Retrieval (IR) system for searching medical literature. Clinicians can easily be overwhelmed by the great number and be inhibited to determine the best possible treatment because hardly do they have enough time to go through each paper they find. It would be much more difficult if a clinician wants to find some papers which exactly match the genetic, disease and demographic information of his patient. This is one of the biggest obstacles on our way towards precision medicine. If physicians could be provided with an effective IR system to easily retrieve relevant medical literature when making decisions, the gap between theoretical setting and real-world clinical setting could be fixed.

For many years, people have been studying how to make medical literature retrieval systems more effective. In the Interactive Information Retrieval research area, according to a survey by Lu \cite{lu2011pubmed}, some approaches have been proposed to improve the retrieval system for better usability, e.g., changing the searching interface which helps users to discover and identify potentially relevant documents. Steinbrook uses the information seeking and information behavior theory to propose better literature searching skills for physicians \cite{steinbrook2006searching}. But these studies do not make great contributions to retrieval systems on the technical level. To promote research for better IR systems in medical literature searching, the National Institute of Standards and Technology (NIST) has organized the Precision Medicine (PM) track since 2017 in the Text Retrieval Conference (TREC). The biomedical abstract task in the PM track uses a real-world setting as discussed above. Suppose a physician is given a patient’s disease, gene and demographic information, how could the IR system help him to retrieve relevant medical literature for conducting treatments?

The problem is very straightforward, but why is it difficult to build an effective IR system for medical literature searching? The first answer is the great volume of articles in the corpus. PubMed\footnote{https://www.ncbi.nlm.nih.gov/pubmed/}, which is a free search engine accessing the MEDLINE database of biomedical topics, has more than 29.1 million records with about 500,000 new records being added each year, according to its Wikipedia page\footnote{https://en.wikipedia.org/wiki/PubMed}. Besides, the vocabulary problem has been found to be a big hurdle, and the vocabulary mismatch between queries and documents is recognized as a common failure in many IR systems. This problem has also been well noted in the health domain \cite{poikonen2009lay}. The vocabulary problem could be interpreted from two aspects. The first aspect comes from obscure medical terminologies and a variety of abbreviations and variations. For example, \emph{BRAF}\footnote{https://en.wikipedia.org/wiki/BRAF\_(gene)} is a human gene which is also referred to as \emph{proto-oncogene B-Raf}, and it has some aliases like \emph{B-RAF1}, \emph{BRAF1} and etc. When a document only includes these aliases instead of the original term, it will not be retrieved if a simple Boolean retrieval strategy is used. The second aspect is that physicians may have varying domain knowledge, which affects how they pose the query. User queries have been found to be very short with only two to three words, and query terms are significant different from terms in professional thesauri \cite{Goeuriot:2017:RSW:3053408.3053422}. If the physician is not very familiar with the domain thesauri, he may have difficulties in formulating a well-structured query. In other words, a physician may input a disease or gene term into the query purely based on his understanding and preferred usage of medical terms, which is probably due to the lack of a unified form for a single concept. Using different terms to describe the same concept could even happen among researchers with comparable level of knowledge on the subject. It is almost impossible for a physician to input all the relevant terms into the query by looking up the term in a thesaurus or an ontology when he attempts to do retrieval. Apart from previous difficulties, even if a physician could formulate a well-structure query and retrieve more relevant documents than he used to, a powerful ranking algorithm is also necessary, considering a physician’s limited time to read only several documents appearing at the top.

In this paper, we propose a new retrieval architecture. We use query expansion to tackle the vocabulary problem by automatically looking up the original disease and gene terms in knowledge bases and adding expansion terms into the query. We aim to retrieve more potentially relevant documents which do not explicitly mention the target disease or gene terms to optimize recall. After this step, we train a Logistic Regression classifier and apply it to the retrieved result to predict the probability of how relevant a document is to the PM track, and we aim to use the probability to re-rank the result to optimize precision. It should be noted that the term classifier and reranker are used interchangeably in this paper. The dataset provided by the PM track which consists of 26.8 million MEDLINE articles and 70,026 ASCO/AACR conference articles serves as our corpus. For training the classifier and tuning parameters, we use the 30 topics from the 2017 track as our training set, and all the retrieval tasks for testing use the 50 topics from the 2018 track.

The paper proceeds as follows. The literature review section provides some preliminary knowledge of the techniques used in this paper and current research in medical literature IR systems as well. The method section introduces how the query is expanded and how the classifier is trained to re-rank the retrieved results. The experiment section shows technical details in implementing the architecture, the comparison between several retrievals with different strategies and several case studies to support why our architecture works. Finally, the conclusion section summarizes all the work in this paper and provides an outlook for the further work.

\section{Literature Review}
In this section, we review relevant literature to provide an overview of the research background and related works mainly from three aspects: query expansion and learning-to-rank in IR systems, applications of knowledge bases in IR systems, and different approaches to improve medical literature IR systems so far.

\subsection{Query Expansion}
Different people may use quite different words to describe a same object. Therefore, traditional IR systems which only use keywords in the query to match terms in documents are not effective in some cases because the information about a same issue can be expressed in different words that may not be exactly the same. A document could still be relevant without the exact terms from the query, as long as it has other words with the same meaning.

When a query contains multiple terms regarding to a user’s desired information, the result could be satisfying with a relatively high recall. However, if a user has poor domain knowledge and the query is short with ambiguous or misspelled words, there might be few relevant documents retrieved. For instance, the example of BRAF in the previous section is common in the medical domain, because many medical terms have aliases and variations and are written in full name or in abbreviation interchangeably. This is the vocabulary mismatch problem which leads to the ineffectiveness of traditional information retrieval systems. It is also where the motivation of query expansion comes.

Query expansion enhances the original query with other words which are most likely to capture user’s desired information. The core idea is to generate an alternative or expanded query for the user. One intuitive way is to add into the query some synonyms or relevant terms of each keyword by looking up in a thesaurus or discovering term relations like their co-occurrences in documents \cite{Manning:2008:IIR:1394399}. Basically, there are two approaches of query expansion: the corpus-based approach which generates expansion terms from documents ranking at the top with the assumption that the most frequent terms appearing in these documents are highly correlated to the original query, and the resource-based approach which leverages external resources like domain-specific dictionaries, ontologies, or knowledge bases to add expansion terms into the original query. In other words, the first approach depends on the searching process and uses relevance feedback in iteration of searching to identify expansion terms, while the second approach is independent of the searching process and additional expansion terms are derived from a knowledge structure.

In the corpus-based expansion approach, relevance feedback is a representative technique that is widely used to improve retrieval performances. The relevance feedback process usually happens after a retrieval is done, and the original query is reformulated according to the retrieval result. Relevance feedback could further be split into two methods which are implicit and explicit. The explicit method asks users to mark documents in the retrieval result as relevant or irrelevant. Then based on the feedback, the algorithm computes a better representation of the information, from which a revised query is formulated and a new retrieval is run. A classical method is the Rocchio algorithm \cite{rocchio71relevance}, which adds an arbitrary percentage of relevant and minuses that of non-relevant documents to the original query in a vector space model. Apart from the explicit method with manual input, Pseudo Relevance Feedback (PRF), also known as blind relevance feedback, is also widely used. It automates the manual part so that the user gets improved retrieval performance without an extended interaction. This approach runs a query to find an initial set of most relevant documents, then makes the assumption that the top K documents are all relevant and revise the query under this assumption.

In the resource-based expansion approach, there are well-structured knowledge resources developed by domain experts in some specific domains, which provide expansion terms based on relations between entries. Expansion terms with certain relations to the original query terms are generated from these knowledge bases.

However, although query expansion helps to capture user’s real information needs, the additional terms may cause the drift of the focus of a search topic caused by improper expansion and may hurt the retrieval performance in both precision and recall. Therefore, apart from generating expansion terms arbitrarily, a lot of studies show that the quality of added terms and weights assigned to these expansion terms could largely affect the result of retrieval. In the medical IR area, Xu \emph{et al.} used PRF to generate expansion terms which are mapped into MeSH, and refined the candidate expansion terms by training term-ranking models to select the most relevant ones for query expansion \cite{8279513}.

\subsection{Learning-to-rank}
The goal of ranking in IR is to generate a ranked list of retrieval results according to their relevance to the original query. Usually a ranking model computes a score between the document and the query to measure their relevance, and the results are ranked in a descending order by the score. Basic IR models like Boolean retrieval could only tell whether a document is relevant or not, but could not tell how relevant a document is. Therefore, the vector space model, language model and other models are introduced to compute the relevance scores. For example, in the vector space model, documents and queries are represented by vectors with TFIDF weights. Then the relevance score is computed by the cosine similarity between two vectors. In the language model, a maximum likelihood method is used to compute the probability to measure the relevance. These ranking models have been widely used in a lot of studies as well as some industry-level applications.

However, one drawback of these traditional ranking models is that there are some parameters which need to be tuned in a heuristic way to get a reasonable result. Apart from the fact that parameter tuning is a time-consuming and trivial task, a model tuned on the validation set sometimes performs very poorly on unseen test queries \cite{INR-016}. Recently years, machine learning has been demonstrating its power in various areas on a lot of tasks. In IR, we could also leverage machine learning techniques on ranking models to do automatically parameter tuning and avoid the overfitting problem.

Adopting the idea of supervised learning, learning-to-rank methods construct the loss function by incorporating the ranking-based information. Same as other supervised learning methods, data are split into training data and test data. Suppose that there is a corpus of documents, and the ranking model is thus trained by a number of queries and each query is associated with a set of retrieved documents with relevance judgments. In testing, given a new query, the ranking function is supposed to create a precise ranked list for retrieval results \cite{Qin:2010:LBC:1842549.1842571}. There are three different types of learning-to-rank methods, which are the pointwise approach, the pairwise approach and the listwise approach. And these three methods model takes a single document, a pair of documents and a list of documents as instances for learning respectively \cite{INR-016}. In training the model, since the document-query are represented in multi-dimensional feature vectors, the selection of features plays an important role regarding to the ranking performance.

\subsection{Medical Information Retrieval}
The quantity of medical articles keeps increasing rapidly, which puts much pressure on medical specialists who need to be aware of up-to-date studies, and physicians who need academic papers as references to diagnose diseases and find treatments. The current retrieval systems for medical literature, however, are not efficient and the searching task is still time-consuming.

There are a lot of studies proposing different methods to improve medical literature searching systems. Lu  did a survey of 28 medical literature searching systems which were built between 1999 to 2010, and sixteen of them were implemented based on the PubMed system with minor adjustments \cite{lu2011pubmed}. These improvements are summarized into four development themes: (1) a new ranking of searching results from the default reverse chronological order to relevance, (2) clustering results into topics for quicker navigation, (3) extracting and displaying semantic relations from results, and (4) improving searching interfaces and retrieval procedures to make the system more user-friendly. However, these improvements were made on the usability level, few changes have been made on the technical level.

In recent years, with the development of technologies and the devotion of human efforts, there are more and more knowledge bases in the medical domain, and these resources are becoming well-structured and comprehensive with a large size. There are some remarkable knowledge bases in the medical domain, e.g., the Medical Subject Headings (MeSH), the Uniﬁed Medical Language System (UMLS), Orphanet and etc. Many studies have successfully incorporated these knowledge bases into IR systems to improve retrieval performances, or to solve the mismatch problem. Martinez \emph{et al.} presented an automatic query expansion method based on random walks over the UMLS graph with applications in the EHR record searching, and showed that query expansion improved the robustness of the system \cite{Martinez:2014:ISO:2953213.2953632}. Otegi \emph{et al.} explored semantic relatedness techniques to propose a generic method using structured knowledge for both query and document expansion to improve results, and showed that their methods are complementary with PRF \cite{Otegi:2015:UKR:2815175.2815228}. Koopman \emph{et al.} proposed a medical IR system that integrates structured knowledge resources, statistical methods, and semantic inference in a Graph Inference retrieval model to tackle the semantic gap problem \cite{Koopman:2016:IRS:2890328.2890349}. Mao \emph{et al.} proposed a new medical IR system which assigns MeSH terms to documents and then quantifies associations between documents and assigned concepts to construct conceptual representations, thus uses the generative concept models match queries and documents \cite{Mao:2015:MDC:2833465.2833486}. Moreover, several studies add knowledge bases into the system to extract latent concepts and relations for query expansion. Griffon \emph{et al.} proposed that terms in documents could be mapped into UMLS to find synonyms to be added into the original queries \cite{PMID:22376010}. Jalali and Borujerdi used the MeSH knowledge base and proposed a query expansion method to match concept pairs between queries and documents. They also pointed out that gene knowledge bases are another source for expansion \cite{4781679}.

Since 2017, the precision medicine track has been added to TREC, and a lot of approaches to improve IR systems for medical literature searching have been proposed by teams participating in the track. Wang \emph{et al.} explored the simplest approach that the NCI thesaurus and COSMIC are used to expand disease and gene terms with synonyms, and the retrieval result proved that the simplest method already significantly improved the retrieval performance \cite{Wang2018MayoNLPTeamAT}. Liu \emph{et al.} explored an iterative approach to construct the query which starts from the strictest situation with exact matching of diseases and genes, i.e., terms must match, to relaxed matching, i.e., some terms may and some terms may not match, and finally to lenient match, i.e., terms may or may not match. The ranked retrieval list is then produced based on the iteration of these queries \cite{medier}. Zheng \emph{et al.} explored assigning different weights for different parts in the query and proposed that the weight for disease, gene, gene mutation and general disease term should be in a descending order \cite{UCAS}. Zhou \emph{et al.} constructed a knowledge graph which contains the relations between gene, cancer and drug pairs from several knowledge bases via knowledge extraction and identifier normalization. Then this knowledge graph serves for both query expansion and re-ranking. A classifier which combines Recurrent Neural Network and Convolutional Neural Network is also trained for re-ranking to retrieved documents \cite{cat}.

\section{Method}
In this section, we first give an overview of our proposed retrieval architecture with two parts of indexing and searching. We then introduce the core techniques to improve retrieval which consists of two consecutive steps: the query expansion step which aims at improving recall and the re-ranking step which aims at improving precision. In the re-ranking section, we discuss how the classifier is trained and how features are generated. Our goal is to retrieve as many relevant documents as possible at the first step, and then re-rank the retrieved results using the reranker to push the most relevant documents to the top. In other words, in the final retrieval result, we expect a higher-precision and higher-recall performance than the baseline strategy.

\subsection{Retrieval Architecture}
\begin{figure}
  \centering
  \includegraphics[width=1.0\linewidth]{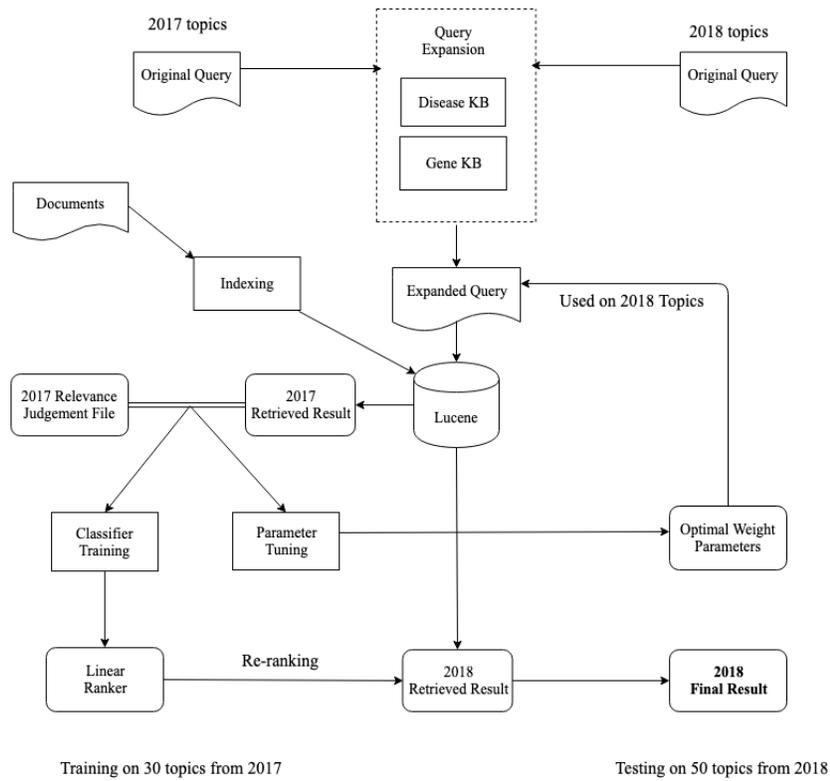}
  \caption{An overview of the retrieval architecture}
  \label{fig:fig1}
\end{figure}

As shown in Figure \ref{fig:fig1}, our novel retrieval architecture consists of three parts: the processing and indexing part which converts the corpus into Lucene indexes, the parameter tuning and classifier training part which uses the 30 topics in 2017 to get optimized weights for expansion terms and a linear classifier for re-ranking, and the final retrieval part to test our proposed strategies which uses the 50 topics in 2018. All the 2017 and 2018 query topics are expanded from the same knowledge bases. We tune and train the model using topics from 2017 and treat the 2018 topics as a black box to ensure the robustness of the model. Finally, we simply apply the same pipeline to the 2018 topics to get a ranked list of documents for relevance judgment.

\subsubsection{Indexing}
We use Apache Lucene\footnote{http://lucene.apache.org/} 6.0.0, which is an open-source, high-performance and publicly-available searching engine toolkit in Java as our main system framework for indexing and searching.

For each paper in the corpus, we index five fields which are document ID, title, content, publication type and MeSH headings (see Table~\ref{index}). It is important to note that the corpus only contain the abstract part of a paper instead of its full text, which is referred as the content in this paper. Only the title and abstract field are analyzed with the built-in standard analyzer, which mean that these two fields are tokenized and used to match the query for relevance measurement. Also, there are 599 documents with slightly different abstract but the same document ID, for a paper is added into the corpus every time after it is revised. We simply index a document at the first hit and ignore the later versions to prevent duplicate document IDs in the retrieved result. Throughout the whole architecture, we use the Okapi BM25\footnote{https://en.wikipedia.org/wiki/Okapi\_BM25} scoring algorithm which has been widely used in various industry-level applications. Compared to the traditional TF-IDF scoring function, BM25 has more reasonable scoring functions and more flexible parameters to be tuned. In this paper, we use the default parameter settings of 1.25 and 0.75 for all the retrieval task.

\begin{table}[!htb]
 \caption{Five fields indexed for each scientific paper}
  \centering
  \begin{tabular}{lll}
    \toprule
    \cmidrule(r){1-2}
    \textbf{Fields} & \textbf{Analyzed} & \textbf{Stored} \\
    \midrule
    Title & True & True     \\
    Abstract & True & True         \\
    ID & False & True         \\
    Publication Type & False & True         \\
    MeSH Heading & False & True         \\
    \bottomrule
  \end{tabular}
  \label{index}
\end{table}

Given a query \emph{Q}, containing keywords $q_1$,...,$q_n$, the BM25 score of a document \emph{D} to the query is calculated as:

\begin{equation}
Score (D, Q) = {\sum _{i=1}^{N} IDF(q_i) * {\frac {f(q_i,D) * (k_1 + 1)} {f(q_i,D) + k_1 * (1-b+b*\frac{|D|}{avgdl})}}}
\end{equation}

where \emph{f($q_i$,D)} is query term $q_i$'s term frequency in the document \emph{D}, \emph{|D|} is the length of document \emph{D} (total number of words), \emph{avgdl} is the average document length in the corpus, \emph{$k_1$} and \emph{b} are hyperparameters. \emph{IDF($q_i$)} is the inverse document frequency (IDF) weight of query term $q_i$, which is usually computed as:

\begin{equation}
IDF (q_i) = log \frac{N - n(q_i) + 0.5}{n(q_i) + 0.5}
\end{equation}

where \emph{N} is the total number of documents in the collection and \emph{n($q_i$)} is the number of documents which contain the query term \emph{$q_i$}.

\subsubsection{Training and testing}
All the training and parameter tuning tasks are done on the 2017 topics, with the released relevance judgement file which contains the true relevance scores of some documents in the corpus. In both training and testing processes, the queries are first expanded using the same knowledge bases which are introduced in the query expansion section, and the initial weight for each expansion term is set to 0.3 to ensure the expansion terms do not shift the focus of the original query too much.

Since the standard number of retrieved documents for each topic is 1000, we use the metric of Recall@1000 to evaluate the retrieval after query expansion, which is calculated as the number of relevant documents in the retrieved 1000 results divided by the total number of relevant documents in the corpus. Then we tune the weight parameters for expansion terms to optimize R@1000, which desires to retrieve as many relevant documents as possible. The results of optimal weight parameters are shown in detail in the experiment section. Another significant contribution of the 2017 topics is to train a linear classifier which predicts the probability of a document’s being relevant to the PM track. After generating high-level features from the text, we train a Logistic Regression model for binary classification in which labels are only relevant or non-relevant. This process is introduced in detail in the re-ranking section. After getting a set of weight parameters optimizing R@1000 and a trained Logistic Regression classifier, we apply the same process on the 2018 topics for testing.

\subsection{Query Expansion}
The goal of query expansion is to use external knowledge bases to expand the original term which helps the query to match more documents which do not contain that term. Since both disease and gene information are given regarding to a patient, we expand both two fields in knowledge bases separately.

\subsubsection{Diseases}
There are a lot of knowledge bases in the general domain, e.g., Wikipedia \footnote{https://www.wikipedia.org/} and DBpedia\footnote{https://wiki.dbpedia.org/}, and Medical Subject Headings (MeSH)\footnote{https://www.ncbi.nlm.nih.gov/mesh} from NCBI in the medical domain. Most research in medical IR systems use the MeSH knowledge base because of its coverage and authority. However, in this paper we use another knowledge base called Lexigram\footnote{https://www.lexigram.io/} which is a publicly-available API. It contains not only the MeSH ontology, but also other knowledge bases like the Systematic Nomenclature of Medicine Clinical Terms (SNOMED CT) and the International Classification of Diseases (ICD). By integrating three professional knowledge bases together, this knowledge base provides broader and more accurate terms for disease expansion. Both MeSH and Lexigram support finding the preferred term, synonyms, as well as parent disease and children disease of a given disease term in a hierarchical structure.

In selecting an appropriate query expansion structure, at first we include all these expansion terms, but we find the parent and children expansion terms bring too many noises into the original query. It makes the retrieved documents either too broad or too specific, and the result is low in both recall and precision. Therefore, in the final query expansion for diseases, parent and children disease terms are discarded. The table below illustrates how expansion terms for a disease term are different in two medical knowledge bases (see Table ~\ref{disease_expansion}).

\begin{table}[!htb]
 \caption{Example of disease expansion terms in two different knowledge bases}
  \centering
  \begin{tabular}{lll}
    \toprule
    \textbf{Knowledge Bases} & \textbf{Lexigram} & \textbf{MeSH} \\
    \textbf{Original Disease Term} & cholangiocarcinoma & cholangiocarcinoma\\
    \textbf{Preferred Term} & cholangiocarcinoma of biliary tract & cholangiocarcinomas\\
    \textbf{Synonyms} & bile duct carcinoma, & Intrahepatic Cholangio-\\
     & bile duct adenocarcinoma, & carcinoma, Extrahepatic\\
     & cholangiocellular carcinoma & Cholangiocarcinoma\\
    \bottomrule
  \end{tabular}
  \label{disease_expansion}
\end{table}

Another edge of using the Lexigram knowledge base is that it can recognize disease terms from input text and automatically generate expansion terms. However, if using the MeSH knowledge base, we need to look up the disease in MeSH first to get a unique identifier, and then retrieve expansion terms according to that identifier. This brings a problem that when the disease in the query topic is a rare disease or do not match the standard disease name in MeSH, the identifier is hard to be found and no results could be retrieved from the ontology.

Apart from the preferred term and synonyms, acronyms are also used as expansion terms in this paper. Consider such a snippet in a paper’s abstract which is relevant to \emph{lung cancer}:

\leftskip1cm\relax
\rightskip1cm\relax
\noindent
“Microarray analyses have revealed significantly elevated expression of the proto-oncogene ROS1 receptor tyrosine kinase in 20-30\% of \textbf{non-small cell lung carcinomas (NSCLC)}. Selective and potent ROS1 kinase inhibitors have recently been developed and oncogenic rearrangement of ROS1 in \textbf{NSCLC} identified. We performed immunohistochemical evaluation of expression of ROS1 kinase and its downstream molecules in 399 \textbf{NSCLC} cases. ROS1 expression in primary and recurring lesions of 92 recurrent \textbf{NSCLC} cases was additionally analyzed.”

\leftskip0cm\relax
\rightskip0cm\relax
In this text, we can clearly see that the disease \emph{non-small cell lung carcinomas} is only written in its full name when it appears in the paragraph for the first time. Then it is written in its acronym form \emph{NSCLC} afterwards for simplicity. This is very common in biomedical scientific papers that people use short acronyms instead of full names because of the length. Therefore, if the query does not include the acronym of a disease, some relevant papers could not be matched or not rank in the top because of the low term frequency of the original disease terms. However, to best of our knowledge, there does not exist a publicly available acronym dictionary or API which automatically maps a disease term to its acronym. For this paper, we simply use such a regular expression \emph{Disease Name ([A-Z]+)} in the corpus to match a pattern that a disease name is followed by a capitalized term in parentheses to retrieve acronyms for each disease.

\subsubsection{Genes}
For query expansion of the gene field, we enrich the original gene term with the genetic dataset provided by the National Center for Biotechnology Information (NCBI)\footnote{ftp://ftp.ncbi.nlm.nih.gov/gene/DATA/}. We simply include the aliases for each gene term by looking it up in the dataset (see Table ~\ref{gene_expansion}).

\begin{table}[!htb]
 \caption{Example of gene expansion terms in the NCBI gene list}
  \centering
  \begin{tabular}{ll}
    \toprule
    \textbf{Knowledge Bases} & NCBI gene list \\
    \textbf{Original Gene Term} & KRAS\\
    \textbf{Aliases} & C-K-RAS|CFC2|K-RAS2A|K-RAS2B|K-RAS4A|K-RAS4B\\
     & K-Ras|KI-RAS|KRAS1|KRAS2|NS|NS3|RALD|c-Ki-ras2\\
    \bottomrule
  \end{tabular}
  \label{gene_expansion}
\end{table}

To summarize, for each query topic the disease term is expanded to its preferred term, acronym and synonyms, and the gene term is expanded to its aliases. This process assists to enrich the information in the original query and thus match those potentially relevant documents which do not mention the original disease or gene term but terms referring to the same disease or gene. In common practice, the expansion terms usually have a lower weight than the original term, in case they introduce too many noises and shift the focus of the original query. Following a heuristic approach, we give the highest priority to the original disease and gene term with a weight of 1, and give a weight between 0 and 1 to decrease the importance of expansion terms. This is supposed to help to improve R@1000 and not to hurt precision at the same time. We then tune these weight parameters on the 2017 topics and the result is in the experiment section.

\subsection{Re-ranking}
Although with query expansion we may optimize R@1000 to retrieve more relevant documents than before, we are still in need of an effective ranking method which pushes the most relevant documents to the top. In the real-world setting, it is impossible for physicians to go through all the 1000 papers retrieved, and their focus should only be on the documents ranking at the top. Therefore, the PM track evaluate the retrieved result by P@10 and R-precision, which calls for a high precision of the retrieval. In this section, we discuss our work of re-ranking to optimize P@10, which is calculated as the number of relevant documents in the top 10 results divided by 10. We present a heuristic approach first, followed by the learning-to-rank approach.

\subsubsection{A Heuristic Approach}
Before starting training the reranker which helps to judge how relevant a document is relevant or how much a document is relevant to the PM topics, we explore a heuristic approach to re-ranking the retrieved results.

First we print out the title and content of retrieved results after the query expansion and examine the top 10 articles for each topic to investigate the most intuitive characteristics of relevant and non-relevant documents. We notice that on the average of 30 topics, almost all the relevant articles have the disease term in the title and most of the non-relevant articles do not. Therefore, we hypothesize that a relevant article should have the disease term in the title and it is an important characteristic, otherwise it shall be penalized to a lower relevance score. Based on this idea, we explore a heuristic approach of punishing those articles without the disease term by multiplying their original scores with a penalty factor between 0 to 1.

To select an approximate penalty factor, we investigate how to push a relevant document which has the disease term in its title and the highest ranking outside the top 10, i.e., ranking at 11, 12, etc., to the top 10. Thus, our goal is to find the gap between such a relevant article and a non-relevant document which does not have the disease term in its title and the highest ranking in the top 10. In other words, we push relevant documents upwards by giving the positions of non-relevant documents.

Finally, we find scores of such relevant documents are around 0.6 of scores of such non-relevant documents on the average of 30 topics. We thus set the penalty factor to 0.6, and the performance of this retrieval shown in the experiment section proves that simple as this heuristic approach is, it does help to push relevant documents upwards and to improve Precision@10.

\subsubsection{Learning to Rank}
The heuristic approach is the starting point of re-ranking the retrieved result to enhance P@10. However, this approach is based on human efforts of reading and examining the results, and such method may not be robust. Therefore, apart from this approach, we aim to leverage the pointwise learning-to-rank idea and machine learning techniques to learn a ranker which can automatically tell whether a paper is relevant to the PM track or not. The feature of whether the disease term appears in the article title shows its usefulness above. Table ~\ref{features} lists query-document features and document-specific features used in the ranker. We choose high-level features instead of pure word features from the bag-of-words model because such features would be too sparse to have a good performance.

First, we hypothesize that an article which is relevant to the topic or in accord with the precision medicine track should focus on treatments and diagnosis of patients, and should not talk about laboratory experiments or some general topics. Second, since each article has the information of its publication types, e.g., journal article, review, case study, clinical trial and etc., we hypothesize that an article with the type of clinical trial should have a higher probability of being relevant. Third, each article has the information of its MeSH headings to imply what its content is mainly about, e.g., Humans, Mutation, genetics, and etc. It is similar to keywords in other scientific papers but uses the MeSH thesaurus. We hypothesize that an article with certain headings should be more likely to be a relevant one.

The features of whether the disease term appears in the title and whether the publication type is clinical trial are categorical, i.e., 0 or 1, and all the other features are numerical because we simply count the term frequency of each keywords.

\begin{table}[!htb]
 \caption{Features in the classifier}
  \centering
  \begin{tabular}{ll}
    \toprule
    \cmidrule(r){1-2}
    \textbf{Feature Description}     & \textbf{Data Type} \\
    \midrule
    Whether the disease name appears in the title & Categorical \\
    How many positive keywords appear in the title & Numerical \\
    How many positive keywords appear in the abstract & Numerical \\
    How many negative keywords appear in the title & Numerical \\
    How many negative keywords appear in the abstract & Numerical \\
    Whether the publication type is about Clinical Trial & Categorical \\
    How many heading keywords appear in the heading & Numerical \\
    \bottomrule
  \end{tabular}
  \label{features}
\end{table}

The next step is to generate the positive keywords, negative keywords and heading keywords for counting. Oleynik \emph{et al.} use the Latent Dirichlet Allocation (LDA) model to do Topic Modeling (TM) among the relevant articles and non-relevant articles to find positive and negative keywords which are strongly correlated to these two corpuses \cite{hpidhc}. Besides the LDA model, we also explore the simple TF-IDF weights of keywords in these two types of corpuses, and the heading keywords are generated in this way as well. These keywords are supposed to have the biggest correlations to the certain corpus. The lists of positive, negative and heading keywords are shown in Table ~\ref{keywords}.

\begin{table}[!htb]
 \caption{Positive, negative and heading keywords}
  \centering
  \begin{tabular}{ll}
    \toprule
    \textbf{Positive Keywords} & treatment, survival, prognostic, clinical, prognosis, therapy \\
     & outcome, resistance, targets, therapeutic, immunotherapy \\
     \midrule
     \textbf{Negative Keywords} & pathogenesis, tumor, development, model, tissue, mouse \\
      &  specific, staining, dna, case, combinations \\
     \midrule
     \textbf{Heading Keywords} & Humans, Mutation, genetics, drug therapy, metabolism \\ 
     & drug therapy, pharmacology, antagonists \& inhibitors \\
     & drug effects, therapeutic use, immunology \\
    \bottomrule
  \end{tabular}
  \label{keywords}
\end{table}

\subsection{Ranker Construction}
We train a Logistic Regression model to predict the probability that a document is relevant to the PM track. All the hyperparameter settings are as default in the Python scikit-learn\footnote{https://scikit-learn.org/} library. For each retrieved article, the feature vector is generated based on the feature table above and the probability of a positive (relevant) label is predicted using the logistic regression model.

We use both the predicted probability of how relevant an article is to the PM track and the original BM25 score for re-ranking, as the experiment result confirms our hypothesis that purely using the probability and ignoring the ranking from original scores do not work very well. However, since the probability is a number between 0 to 1 and the original BM25 score could be as large as 70, we should keep them at the same scale. Therefore, we do the min-max transformation on the original BM25 score and it is thus scaled to a number between 0 to 1, which makes it more reasonable to add the probability to the original score. Then for each article, a new score is generated by adding up the raw BM25 score and the probability from the classifier, finally the re-ranking is based on new scores.

However, in experiments we find that applying the reranker to all the 1000 results do not improve the precision, but hurt the performance instead. We suppose that it is because the positive instances in the training set merely come from past teams’ top retrieved results, which are much fewer than the negative instances. In other words, since there are 26.8 million articles in the corpus, it is impossible for people to manually label each document as relevant or non-relevant. Thus, the ad-hoc relevance judgment by experts are only applied on documents which rank at top in each team’s submitted results, which leads to a different distribution of instances in the training and testing set. Therefore, instead of applying the reranker on all the 1000 retrieved documents to do re-ranking, we only re-rank the top 50 documents, after testing different top K documents to re-rank by optimizing P@10 on the 2017 topics.

In summary, we train the reranker on the 2017 topics and find that the optimal number to re-rank is 50. When we run queries on the 2018 topics, we follow the heuristic approach first, i.e., we punish a document which does not mention the disease term in its title by multiplying its raw score by a penalty factor of 0.6, based on which the results are re-ranked for the first time. We then transform the raw BM25 scores using a min-max scaler to a number between 0 and 1. Afterwards, we only input the top 50 documents into the reranker and add up BM25 scores with the probability, and keep scores of the other 950 documents unchanged. Based on new scores, the re-ranking is done for the second time and a final ranked list is returned for relevance judgment.

\section{Experiment}
In this section, we discuss the performances of different retrieval strategies and methods. We first introduce the dataset which is used in this paper, and then show how a query is generated for each topic. We then compare retrieval performances in the result section followed by three case studies and some other approaches which fail to show improvements but may work in the future.

\subsection{Dataset Overview}
As is discussed in previous sections, we use the dataset which consists of 26.8 million MEDLINE articles and 70,026 ASCO/AACR conference articles as corpus, and all these articles only contain the abstract part instead of the full text which is referred to as the content field in this paper. The indexing process takes about 8 hours on a research computing node with 16 core CPU and 100 gigabytes RAM, and the file size of indexes is about 18 gigabytes.

\begin{figure}[!htb]
  \centering
  \includegraphics[width=0.7\columnwidth]{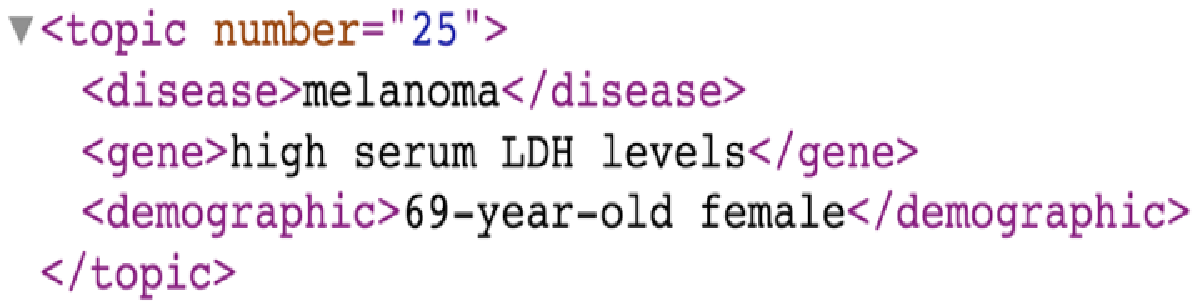}
  \caption{A query topic}
  \label{fig:data1}
\end{figure}

There are 30 topics in the 2017 track and 50 topics in the 2018 track, both are in the XML format and each node represents a patient with disease, gene and demographic information (see Figure \ref{fig:data1}).

\subsection{Query Structure}
For each topic, we generate the query using the disease and gene information, and the demographic information is not used in this paper. After these two fields are expanded in knowledge bases, weights are assigned to each term. We use the OR operator between all the terms, because it is impossible for a paper to contain all the query terms after expansion. We use the SHOULD operator for the title field and the MUST operator for the content field, which means it is not required for a document to match the query in its title but required to match the query in its content.

\subsection{Result and Implications}
Our final query structure and weights for each expansion part are shown in Table ~\ref{weights}. We keep weights of the original disease and gene terms always as 1, and we set weights of all the expansion terms to 0.3 initially. To optimize Recall@1000, we control the weight of other expansion terms in the query as the same and tune one part to an optimized recall.

The final weights are in accord with our expectations. A disease has 3 or 4 preferred terms and synonyms in total, and these terms usually contain several words (see Table ~\ref{disease_expansion}). Also, when we examine the output, we find that some expansion terms may not appear in the whole corpus because they are too rare, or not be relevant to the original disease. Therefore, the weight of these expansion terms should not be too high to shift the focus of the original query and it is thus set to 0.1. However, the weight of disease acronyms is 0.5 because it is highly related or even equivalent to the original disease. The weight for gene aliases is fairly 0.3.

\begin{table}[!htb]
 \caption{Weights for each term which optimizes R@1000}
  \centering
  \begin{tabular}{lll}
    \toprule
    \textbf{Field} &  \textbf{Query Term} & \textbf{Weight} \\
    \midrule
    \textbf{Diseases} & Original Disease Term & 1 \\
    & Disease Preferred Term & 0.1 \\
    & Disease Synonyms & 0.1 \\
    & Disease Acronyms & 0.5 \\
    \textbf{Genes} & Original Gene Term & 1 \\
    & Gene Aliases & 0.3 \\
    \bottomrule
  \end{tabular}
  \label{weights}
\end{table}

\begin{table}[!htb]
 \caption{Comparison between retrieval strategies}
  \centering
  \begin{tabular}{llll}
    \toprule
    \textbf{Retrieval Strategy} & \textbf{R@1000} & \textbf{P@10}  & \textbf{R-prec} \\
    \midrule
    \textbf{Baseline} & 0.686 & 0.544  & 0.317 \\
    \textbf{Retrieval I} & 0.695 & 0.554  & 0.299 \\
    \textbf{Retrieval II} & 0.702 & 0.568  & 0.320 \\
    \textbf{Retrieval II} & 0.702 & 0.618  & 0.322 \\
    \textbf{Retrieval IV} & 0.702 & \textbf{0.646}  & \textbf{0.336} \\
    \midrule
    \textbf{hpi-dhc} & - & \textbf{0.706}  & \textbf{0.365} \\
    \textbf{Cat Garfield} & - & 0.668 & 0.325 \\
    \textbf{SIBTextMining} & - & 0.632 & 0.357 \\
    \bottomrule
  \end{tabular}
  \label{comparison_retrieval}
\end{table}

The comparison between different retrieval strategies is shown in Table ~\ref{comparison_retrieval}. We run five retrievals in total and we describe the settings for each retrieval below. To be more specific, \textbf{Retrieval I \& II} are used to test the hypothesis that query expansion is useful, and these two retrievals are different in the structure of expansion terms. And based on \textbf{Retrieval II} which optimizes Recall@1000, \textbf{Retrieval III \& IV} are used to test the hypothesis that re-ranking is effective, and these two retrievals are different in the re-ranking method, with the heuristic and learning-to-rank approach separately. In \textbf{Baseline} and \textbf{Retrieval I, II, III \& IV} strategies, a same setting is used that terms are connected using the OR operator, and the operators for the title and content field are SHOULD and MUST separately.

\textbf{Baseline}: This serves as the most basic strategy in our approaches. In this retrieval, we use only the original disease and gene terms in the topic without query expansion and re-ranking.

\textbf{Retrieval I}: In this retrieval, the disease term is expanded to its preferred term and synonyms, and the gene term is expanded to its aliases, with weights assigned shown in Table ~\ref{weights}.

\textbf{Retrieval II}: Compared to previous retrieval, acronyms of diseases are added into the expansion terms, with weights assigned shown in Table ~\ref{weights}.

\textbf{Retrieval III}: Compared to \textbf{Retrieval II}, the re-ranking step is added after the first retrieval with the heuristic approach.

\textbf{Retrieval IV}: Compared to \textbf{Retrieval II}, the re-ranking step is added after the first retrieval with the trained reranker.

In summary, \textbf{Retrieval I \& II} prove that adding expansion terms into the original query for diseases and genes is helpful, as they improve Recall@1000 from 0.686 to 0.695 and 0.702 separately, which means more potentially relevant documents are being matched because of these expansion terms. The higher recall in \textbf{Retrieval II} over that in \textbf{Retrieval I} illustrates that acronyms are important in expansion. Also, with a wise selection of weight parameters in the query expansion step, P@10 is not hurt and, instead, improved from 0.544 to 0.554 and 0.568 separately. This result supports that high-quality expansion terms could help to improve both recall and precision.

Based on \textbf{Retrieval II} which optimized Recall@1000, \textbf{Retrieval III \& IV} prove that the re-ranking step improves precision by pushing the most relevant documents to the top. \textbf{Retrieval III} shows that the heuristic approach which penalizes documents without disease term in the title already works well, as P@10 increases from 0.568 to 0.618, considering its simplicity. \textbf{Retrieval IV} proves that a Logistic Regression reranker further helps to improve the precision, as the precision is 0.1 higher than the baseline approach and 0.026 higher than the heuristic approach.

In Table ~\ref{comparison_retrieval}, we include retrieval performances from three top-ranking teams in the 2018 PM track as well. Team \emph{hpi-dhc}, \emph{Cat Garfield} and \emph{SIBTextMining} have the highest P@10 among all the teams. The extreme high precision of team \emph{hpi-dhc} and \emph{Cat Garfield} benefits from their domain knowledge. Team \emph{hpi-dhc} make hand-crafted rules for documents matching in retrieval, and they pre-process the corpus in their own framework, which is an edge over our practice of indexing the raw text \cite{hpidhc}. Team \emph{Cat Garfield} construct their own knowledge graph by integrating several knowledge bases together, and this high-quality knowledge graph proves its power in both query expansion and re-ranking. Also, they train a CNN classifier for re-ranking, which is more complicated but has a better performance than our linear classifier \cite{cat}. Unfortunately, team \emph{SIBTextMinining} does not provide a report of their work. It is highlighted that with query expansion in a publicly available knowledge base and a linear classifier for re-ranking, the Precision@10 in this paper already ranks at the third place among all the teams.

\subsection{Case Studies}
To further demonstrate how query expansion and re-ranking help to improve the retrieval performance clearly, in this section we present case studies with examples to illustrate how these two steps enhance retrieval in \textbf{Case I \& II}. We also give an error analysis case to discuss why our retrieval architecture fails to handle some tough and ambiguous situations in \textbf{Case III}. Considering the length of full texts, we only provide snippets in this section and full abstracts could be found at the Appendix.

\textbf{Case I}: The contribution of query expansion.

As is discussed in previous sections, the motivation of query expansion is to enrich query information by adding into the query alternative or relevant terms of the original keywords, and it is expected to match some potentially relevant documents which do not contain those original keywords. Hereby we show the contribution of query expansion with an article which is not retrieved in the baseline model but is retrieved after the query expansion technique is used.

\begin{figure}[!htb]
  \centering
  \includegraphics[width=0.7\columnwidth]{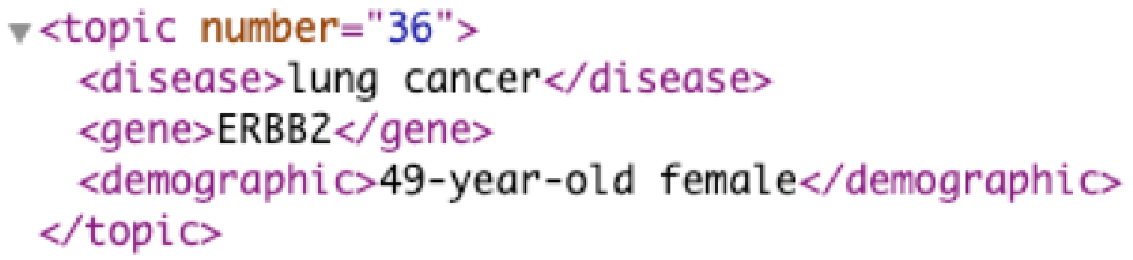}
  \caption{A query topic}
  \label{fig:data2}
\end{figure}

As is shown in Figure ~\ref{fig:data2}, topic No.36 aims to find relevant articles about \emph{lung cancer} and the \emph{ERBB2} gene. And below is the snippet of one paper’s abstract (PubMed ID: 14981584) which is relevant to this topic. It demonstrates how expanding disease and gene terms helps to retrieve more relevant documents.

\leftskip1cm\relax
\rightskip1cm\relax
\noindent
“Title: The role of \textbf{HER-2/neu} expression and trastuzumab in non-small cell lung cancer.

\noindent
Abstract: ... The \textbf{HER-2/neu} receptor... \textbf{HER-2/neu} is overexpressed in 16\% to 57\% of patients with \textbf{non-small cell lung cancer (NSCLC)} and studies have shown that \textbf{HER-2/neu} overexpression imparts a poor prognosis in both resected and advanced \textbf{NSCLC}, as it does in ... the \textbf{HER-2/neu} protein receptor, has been ... with \textbf{HER-2/neu}-positive metastatic breast cancer. In \textbf{NSCLC} preclinical studies, marked synergistic ... in \textbf{HER-2/neu}-expressing cell lines. However, to date, clinical studies with trastuzumab in patients with \textbf{NSCLC} have not shown a demonstrable advantage ...”

\leftskip0cm\relax
\rightskip0cm\relax
As we can see from this example, this paper does not talk about the \emph{ERBB2} gene at all, but it mentions \emph{HER-2/neu} which is an alias of \emph{ERBB2} throughout the abstract. This paper is not retrieved in the baseline strategy because of the mismatch between the original gene term and gene terms in the content. However, it is retrieved after the gene term is expanded to its aliases, i.e., the expansion term \emph{HER-2/neu} is included into the query. Also, this paper does not talk about the original disease term lung cancer, but one of its synonyms which is \emph{non-small cell lung cancer} and its acronym \emph{NSCLC} are used five times in total merely in the snippet above. If the disease is not expanded to its preferred term, synonyms and acronym, this paper would not be retrieved because of the low term frequency of the disease term lung cancer. By query expansion, recall@1000 for this topic is thus improved from 0.19 in the baseline approach to 0.48.

\textbf{Case II}: The contribution of re-ranking.

As is discussed in previous sections, the motivation of re-ranking is to push a relevant document with a low raw score and low ranking to the top, and also to lower the rankings of non-relevant documents. It is expected to increase precision at the top, e.g., Precision@10 used in this paper. In this case, we show the contribution of the re-ranking step with one example that is not in the top 10 results initially but does appear in the top 10 list after re-ranking, and with another example which is punished to a rank after 10. We use the same topic as in the previous case, i.e., \emph{lung cancer} and the \emph{ERBB2} gene. It shall be highlighted that the original Precision@10 for topic No.36 is 0.2, and the new Precision@10 for this topic is increased to 0.5 after re-ranking.

In the re-ranking section, we propose two methods which are a heuristic approach and a learning-to-rank approach, and examples are given for both methods.  We first show how the heuristic approach decreases scores of non-relevant documents which do not mention the disease term in the title. There are 8 non-relevant documents in the initial top 10 list, and none of them appears in the new list after re-ranking. For the sake of simplicity, we give only one example in Table ~\ref{punishing}. 

\begin{table}[!htb]
 \caption{An example of punishing non-relevant documents}
  \centering
  \begin{tabular}{lllll}
    \toprule
    \textbf{Document Title} & \textbf{Raw Score} & \textbf{New Score}  & \textbf{Raw Rank} & \textbf{New Rank}\\
    \midrule
    HER-2/neu and topoisomerase  & 61.8 & 37.08  & 4 & 47 \\
    IIalpha in breast cancer \\
    \bottomrule
  \end{tabular}
  \label{punishing}
\end{table}

As is shown in Table ~\ref{punishing}, a top-ranking article (PubMed ID: 12755489) is now going down to 47 although it ranks as high as 4 initially, because the topic disease \emph{lung cancer} does not appear in its title.

It is not hard to tell from the title that this article is not about \emph{lung cancer} at all, it should all about \emph{breast cancer} instead. But why could this obviously non-relevant article be retrieved? From the snippet of its abstract below, it is revealed that the high-ranking raw score purely comes from the extremely high term frequency of the gene \emph{HER-2/neu}. Simple as the heuristic approach is, it still helps to refine the top results.

\leftskip1cm\relax
\rightskip1cm\relax
\noindent
“Abstract: ... The \textbf{HER-2} (also known as \textbf{ErbB2/c-erbB2/HER-2/neu}) oncogene is the most frequently ... \textbf{HER-2} is also a target for ... \textbf{HER-2} receptor. \textbf{HER-2} has also been implicated ... by the \textbf{HER-2} ... of the \textbf{HER-2} amplified primary breast tumors ... of \textbf{HER-2} amplification also ... \textbf{HER-2} is an oncogene that ... other than \textbf{HER-2}, such as ... \textbf{HER-2} status still ... importance of \textbf{HER-2} ...”

\leftskip0cm\relax
\rightskip0cm\relax
The heuristic approach proves its usefulness in punishing non-relevant documents above, and we give another example of how a relevant document (PubMed ID: 15312350) is boosted to top 10 after re-ranking using the learning-to-rank approach. As is shown below, this document is not punished because it mentions the topic disease \emph{lung cancer} in its title. Also, from the snippet of its abstract, we could see that it is boosted because of the high term frequency of positive keywords \emph{prognostic} and \emph{survival} in both its title and content, which indicates that this paper focus on treatments and is in accord with the PM track. Its actual rankings go with our expectations: its raw score ranks at 16; its probability ranks at 6 and its final score after adding two up ranks at 7.

\leftskip1cm\relax
\rightskip1cm\relax
\noindent
“Title: \textbf{Prognostic} value of expression of FASE, HER-2/neu, bcl-2 and p53 in stage I non-small cell \textbf{lung cancer}

\noindent
Abstract: To evaluate the \textbf{prognostic} value of expression of fatty acid synthase (FASE), .... The 5-year \textbf{survival} rate was lower in HER-2/neu and FASE positive patients than in negative patients, which showed that HER-2/neu and FASE expression were associated with significantly poor \textbf{survival} ... with a 5-year \textbf{survival} rate of 78.2\% ... \textbf{prognostic} factors for \textbf{survival}. HER-2/neu and FASE are independent \textbf{prognostic} factor in stage I non-small cell lung cancer ...”

\leftskip0cm\relax
\rightskip0cm\relax

\textbf{Case III}: Error Analysis.
Although retrieval results in two previous sections have proven that both query expansion and re-ranking are useful in improving the retrieval performance, there are two situations that our proposed retrieval architecture fail to handle. One is that a non-relevant document is retrieved and ranked among the top 10 list, and another is that a relevant document is not retrieved or does not rank in the top 10 list, which could be recognized as cases of false positive and false negative, respectively. Hereby we do an error analysis of these two situations with two examples, from which we investigate the flaw of our retrieval architecture for future improvements. And we still use topic No.36 which is about \emph{lung cancer} and the \emph{ERBB2} gene as above.

We first take a look at one example of the false positive cases, which means that the document below (PubMed ID: 22730705) is non-relevant to the topic but still ranks among the final top 10 results. It ranks at 39 if sorted by the raw score because of its fairly high term frequency of both disease and gene terms. However, it ranks at 5 if sorted by the probability from the reranker, for there are 8 positive keywords in total and no negative keywords at all, which could be seen from the snippet below. Also, the disease term appearing in the title ensures that it is not punished. Since the raw is transformed in a min-max scale, the raw score gap between this document and other documents is shortened, and it finally ranks at 8 in the final result. A close look at this article reveals that it focuses on predicting cancer patients’ survival rather than developing or evaluating cancer treatments. This makes it non-relevant to the central goal of PM track.

\leftskip1cm\relax
\rightskip1cm\relax
\noindent
“Title: \textbf{HER-2/neu} oncogene and estrogen receptor expression in non small cell lung cancer patients

\noindent
Abstract: ... The \textbf{prognosis} is usually ... parameters and \textbf{clinical} stage, but additional \textbf{prognosis survival} factors ... of \textbf{HER-2/neu} and estrogen receptors in \textbf{nonsmall cell lung cancers} and their relation to \textbf{survival} of patients with \textbf{non-small cell lung cancers} and to traditional \textbf{prognostic} factors ... resected patient tissues of \textbf{non-small cell lung cancers}, and the following parameters were examined: \textbf{HER-2/neu} and estrogen receptor expression, as well as the related \textbf{clinical} and pathological features ... Our findings indicate that the expression of \textbf{HER-2/neu} ... the \textbf{survival} of patients with \textbf{non-small cell lung cancers}.”

\leftskip0cm\relax
\rightskip0cm\relax

Apart from the false positive situation, we give another example of the false negative cases, which means that the document below (PubMed ID: 11153605) is relevant to the topic but does not rank among the final top 10 results. Its raw score ranks at 9, which is much higher than the non-relevant document above. It is not only because the high term frequency of disease and gene terms, but more importantly, the gene mentioned in this paper is the original term \emph{ERBB2}, which has a higher weight than the expansion term \emph{HER-2/neu}, i.e., the weight 1 comparing to the weight 0.3. However, it ranks at 14 if sorted by the probability from the reranker, for there are only 3 positive keywords and also two negative keywords case. Therefore, after adding the raw score and the probability up, this article ranks at 13 in the final result.

\leftskip1cm\relax
\rightskip1cm\relax
\noindent
“ Title: Ploidy, expression of erbB1, \textbf{erbB2}, P53 and amplification of erbB1, \textbf{erbB2} and erbB3 in non-small cell \textbf{lung cancer}.

\noindent
Abstract: ... assess the \textbf{prognostic} value of ... \textbf{erbB2} ... in \textbf{non-small cell lung cancer (NSCLC)}. Consecutive patients with \textbf{NSCLC} who underwent \textbf{treatment}... In 108 \textbf{cases}, ... In another 108 \textbf{cases}, ... determined in the tumours of 53 patients ...18\% for \textbf{erbB2} ... 94\% for \textbf{erbB2} ... \textbf{erbB2} and P53 expression ... stage ... \textbf{erbB2} ... not \textbf{prognostic} parameters in \textbf{non-small cell lung cancer} ...”

\leftskip0cm\relax
\rightskip0cm\relax
From these two errors, together with the retrieval compassion in Table VII, we can see that although adding the probability helps to improve precision, it could still make mistakes, especially when a non-relevant document has more positive keywords, or a relevant document has more negative keywords. This implies that the features of simply counting positive or negative keywords in title or abstract are not robust or may not work in some cases. Also, these two errors also suggest that simply adding the raw BM25 score and the probability may not work, as we can see the non-relevant document ranks higher largely because of its high probability, and the relevant document ranks lower because of low probability. Therefore, a more reasonable weighting should be investigated.

\subsection{Other Trials}
Apart from the retrievals discussed above which give the expected performances, we also try other approaches both in the query expansion step and the re-ranking step. Unfortunately, all these approaches do not show improvements of the retrieval result, some of which even lead to a worse performance than the baseline approach. Hereby we present these trials with the intuition behind and error analysis, which could be used for future work references.

\textbf{Trail I}: We use other knowledge bases like MeSH, Orphanet\footnote{https://www.orpha.net/consor/cgi-bin/index.php}, Disease Ontology\footnote{http://disease-ontology.org/}, and Wikidata\footnote{https://www.wikidata.org/wiki/Wikidata:Main\_Page} for disease query expansion. However, diseases terms expanded by these knowledge bases either are too far from the original disease to be relevant, i.e., some rare diseases or terms which are abandoned and not used anymore, or have only minor changes in the order of words or simply to the plural form, e.g., Colonic Neoplasms in MeSH has the preferred term of Neoplasm, Colonic. The result of adding these low-quality expansion terms shows that R@1000 is slightly improved and P@10 is hurt because all the disease expansion terms contain almost same words, which lead to an implicit boost of these duplicate words and shift the focus of the original query. Even though we remove duplicate words in the expanded query by using unique terms, the result still shows that R@1000 is still slightly improved. Therefore, we choose the Lexigram knowledge base for query expansion.

\textbf{Trail II}: We add parent, children and siblings of a disease as expansion terms when using the MeSH knowledge base, for MeSH is an ontology in a tree structure. We hypothesize that some relevant articles may not talk about the disease, but about the parent of the disease in the tree which are broader or the children which are more specific. We even consider the siblings of a disease in the tree structure. However, even we do a lot of work on parameter tuning, the results show that adding these expansion terms hurts both R@1000 and P@10. It is probably because these expansion terms are too broad or specific to be relevant, which adds noises to the query. We then hypothesize that the presence of parent, children or sibling diseases could be feature in re-ranking, and we generate the feature by counting how many these terms occur in the text. However, the result proves that the reranker without this feature works better. After reading about 30 relevant articles, we found only 2 of them contain these parent or children disease terms.

\textbf{Trail III}: We change the operator for the title field from SHOULD to MUST. We hypothesize that setting a strict matching criterion that the title field must contain the terms in the query would improve the precision at one-time retrieval and skip the re-ranking part. However, the results show that both P@10 are not improved, and because of the strict criterion, there are much fewer results retrieved.

\textbf{Trail IV}: We boost the title field with weight 2, comparing to the content field having weight 1 without boosting. We hypothesize that the title field should be more important than the content filed, with the intuition that if an article mentions the disease or gene terms in the title, it might be more relevant to the topic. However, the results show that it hurts both R@1000 and P@10. When reading some articles in the corpus, we find some irrelevant articles also contain the disease or gene terms in the title, and these irrelevant articles are pushed to the top after we boost the title field.

\textbf{Trail V}: We return the abstract text in the retrieved result and then use the Pubtator\footnote{https://www.ncbi.nlm.nih.gov/CBBresearch/Lu/Demo/PubTator/} application to do Named Entity Recognition in text to recognize disease and gene terms. Inspired by the bag-of-words model, we use the bag-of-concepts model along with the PRF method to do re-ranking by running a query in the retrieved articles. The vector space model and the TF-IDF weighting representation of concepts are used in this step.

We first construct the corpus of all the diseases and genes, then each article is represented by a vector containing disease and gene terms. We then generate the query vector for re-ranking by adding the average vector of top K articles to the vector of the original disease and gene terms. We compute the cosine similarity between the query vector and other article vectors, and the articles are ranked according to their similarities to the query. However, it is a time-consuming step because there are several parameters to be tuned, which are how many top ranked documents should be used for averaging, what weights of the original query vector and the averaged query vector should be when adding them up. Although we spend a lot of time tuning these parameters, the optimal results in this strategy are still not as good as expected, which shows only a slight improvement of P@10.

\textbf{Trail VI}: We add positive and negative keywords in the query and give these terms a boost or a penalty. We hypothesize that articles which focus on treatment, clinical study and patients should be more relevant that articles which focus on laboratory experiments. Therefore, we added some keywords suggest by Oleynik \emph{et al.} \cite{hpidhc} into the query to boost documents with positive keywords and penalize documents with negative keywords. However, Lucene does not support negative boost for a query term. Therefore, we use these keywords as our features to train the Logistic Regression classifier in the re-ranking step.

\textbf{Trail VII}: We leverage the disease-gene connections for query expansion and add other variations of a gene. Although the Lexigram knowledge base and the gene database has proved that it works well in the query expansion step and provide comprehensive expansion terms, the expansion is still based on the disease or gene itself. In other words, we are still starting from a disease or a gene and investigating what relevant terms are, e.g., preferred terms and synonyms.

However, we hypothesize that some totally different diseases could be relevant if they are caused by similar genes, and some totally different genes could be relevant if they cause similar diseases, which means we are digging into the latent relations. Many teams have successfully incorporated the knowledge bases of disease-gene-drug relations into the query expansion. Inspired by this idea, we use the DisGeNet\footnote{http://www.disgenet.org/home/} knowledge base which contains relations between diseases and genes and transform the knowledge graph to a matrix of diseases and genes with each element in the matrix representing the probability that a gene and a disease are correlated. Thus, each disease and gene could be represented by a vector and pairwise cosine similarities between vectors could be calculated.

Given a disease or a gene, the relevant terms are generated by sorting the similarities in a descending order. Then we add the top K relevant terms in the query as expansion terms. However, the results show that it does not help to improve R@1000. After examining the expansion terms, we find most expansion terms generated which even rank at the top 5 still have poor similarities, which means they share few diseases or genes information in common. We view this as a potential approach to generate expansion terms, or could be used in the re-ranking step in the future work.

\section{Conclusions}
Medical literature search has long been suffering from the problems of vocabulary mismatch between queries and documents and ineffective ranking methods as well, which makes the traditional Boolean retrieval deliver both low recall and precision. In this paper, we use external knowledge bases in the medical domain to expand the query automatically, which enriches the information in the original query without manual input. This helps the query to match documents without terms from the original query. After the first retrieval, we use a ranker which is trained on the 2017 track topics to predict the probability of how relevant the retrieved document is. Then the retrieved results are ranked by the probability. The experiment results prove that both these two steps help to improve the retrieval performance, and the final model outperforms the baseline model in both recall and precision. The future work could be in the directions discussed below. 
First of all, although the Lexigram knowledge base provides well rounded expansion terms for diseases and outperforms the MeSH ontology, it is possible there exists a knowledge base which provides higher-quality expansion terms. And as is discussed in the literature review section, there has been some studies which focus on the selection of expansion terms instead of adding all of them. The refinement of query expansion terms could be further studied in the future.

Secondly, since the reranker is trained on the 2017 topics, it could be trained again on both 2017 and 2018 topics together when it is applied to the 2019 PM track. With more training data and better feature generation and selection, we expect the reranker to have a better performance and avoid mistakes in the case study section, which is expected to improve the precision further. Also, in training the reranker we are on the level of the whole corpus, and predicting whether an article is relevant to the general PM track rather than to the topic. We hypothesize that there are some features representing the connection between the article and the topic, which makes the classification more specific to each topic. And the case study of error analysis reminds us that the new scoring schema is also a direction of improvement, which could be changed from simply adding the probability and the raw BM25 score to a weighted sum because either the raw score or the probability may be more important than another, or merging two ranks could also work.

Third, as there are both gene and disease information in the topics, we could leverage the connections between diseases and genes (e.g. how strongly a disease and a gene are connected to each other) for relevance judgment by using a knowledge graph. This is a failed trial in this paper, but it still gives us some inspirations.

Last but not least, we use high-level features with a linear model in training the reranker, and it is possible that the deep learning approach could also work. For example, we could train word vectors for a deep Neural Network classifier.

\section{Appendix}

\textbf{Case I: A paper which is retrieved after query expansion}

PubMed ID: 14981584

Title: The role of HER2/neu expression and trastuzumab in non-small cell lung cancer

Abstract: Research over the past decade has led to an increased understanding of the pathophysiology of lung cancer. The HER2/neu receptor is a member of the ErbB family of signaling-transduction receptors and appears to play a major role in the development of lung cancer as well as many other solid tumors. HER2/neu is overexpressed in 16\% to 57\% of patients with non-small cell lung cancer (NSCLC) and studies have shown that HER2/neu overexpression imparts a poor prognosis in both resected and advanced NSCLC, as it does in breast cancer. Trastuzumab, a humanized monoclonal antibody that recognizes the HER2/neu protein receptor, has been approved by the US Food and Drug Administration for patients with HER2/neu-positive metastatic breast cancer. In NSCLC preclinical studies, marked synergistic growth inhibition occurred when standard cytotoxic chemotherapy was combined with trastuzumab in HER2/neu-expressing cell lines. In the clinical setting, trastuzumab has proven safe and feasible in combination with cytotoxic chemotherapy in both single-institution and multi-institutional cooperative group studies. Unlike the experience in advanced breast cancer, cardiac toxicity is a marginal concern in this population. However, to date, clinical studies with trastuzumab in patients with NSCLC have not shown a demonstrable advantage for the majority of patients.

\noindent
\textbf{Case II: A non-relevant paper which is punished in re-ranking}

PubMed ID: 12755489

Title: HER-2/neu and topoisomerase IIalpha in breast cancer

Abstract: In breast cancer, the predominant genetic mechanism for oncogene activation is through an amplification of a gene. The HER-2 (also known as ErbB2/c-erbB2/HER-2/neu) oncogene is the most frequently amplified oncogene in breast cancer, and its overexpression is associated with poor clinical outcome. In addition to its important role in breast cancer growth and progression, HER-2 is also a target for a new form of chemotherapy. Breast cancer patients have been treated with considerable success since 1998 with trastuzumab, a recombinant antibody designed to block signaling through HER-2 receptor. HER-2 has also been implicated in altering the chemosensitivity of breast cancer cells to different forms of conventional cytotoxic chemotherapy, particularly of topoII-inhibitors (e.g., anthracyclines). Topoisomerase IIalpha gene is located just by the HER-2 oncogene at the chromosome 17q12-q21 and is amplified or deleted in almost 90\% of the HER-2 amplified primary breast tumors. Recent data suggests that amplification and deletion of topoisomerase IIalpha may account for both relative chemosensitivity and resistance to anthracycline therapy, depending on the specific genetic defect at the topoIIalpha locus. Expanding our understanding of HER-2 amplification also changes its role in the pathogenesis of breast cancer. HER-2 is an oncogene that clearly can drive tumor induction and growth and is also a target for a new kind of chemotherapy, but its function as a marker for chemoselection may be due to associated genetic changes, of which topoisomerase IIalpha is a good example. Moreover, despite potential evidence that genes other than HER-2, such as topoisomerase IIalpha, may be more important predictors of therapeutic response in breast cancer, HER-2 status still has a very significant role in therapeutic selection, mainly as the major criterion for administering trastuzumab in treating breast cancer. Thus, the clinical and therapeutic importance of the HER-2 and topoisomerase IIalpha status to breast cancer management should only increase in the next few years.

\noindent
\textbf{Case II: A relevant paper which is boosted in re-ranking}

PubMed ID: 12755489

Title: Prognostic value of expression of FASE, HER-2/neu, bcl-2 and p53 in stage I non-small cell lung cancer

Abstract: To evaluate the prognostic value of expression of fatty acid synthase (FASE), HER-2/neu, bcl-2 and p53 in stage I non-small cell lung cancer (NSCLC). Expression of FASE, HER-2/neu, bcl-2 and p53 protein was detected by immunohistochemical staining in 84 patients with stage I NSCLC who underwent surgery. Multiple clinical parameters and survival were analyzed. The expression of FASE, HER-2/neu, bcl-2 and p53 was 29.8\%, 40.5\%, 33.3\% and 39.3\%, respectively. The local recurrence and bone-metastasis rate were higher in FASE positive patients than in negative patients (28.0\% vs 10.2\%, P = 0.05; 61.5\% vs 23.9\%, P = 0.017, respectively). The 5-year survival rate was lower in HER-2/neu and FASE positive patients than in negative patients (37.7\% vs 67.7\%, P = 0.0083; 35.1\% vs 66.1\%, P = 0.0079, respectively), which showed that HER-2/neu and FASE expression were associated with significantly poor survival. Patients whose tumors were both HER-2/neu and FASE negative had better outcome, with a 5-year survival rate of 78.2\%, compared with 36.3\% in those whose tumors were positive for either one (P = 0.002). However, bcl-2 and p53 were not independent prognostic factors for survival. HER-2/neu and FASE are independent prognostic factor in stage I non-small cell lung cancer patients who expressed one or both markers.

\noindent
\textbf{Case III: A non-relevant paper which ranks among top 10}

PubMed ID: 22730705

Title: HER-2/neu oncogene and estrogen receptor expression in non small cell lung cancer patients.

Abstract: Non-small cell lung cancers are among the leading causes of cancer morbidity and mortality worldwide. The prognosis is usually based on traditional pathohistological parameters and clinical stage, but additional prognostic survival factors have also been sought. The aim of this retrospective study was to explore the membranous expression of HER-2/neu and estrogen receptors in nonsmall cell lung cancers and their relation to survival of patients with non-small cell lung cancers and to traditional prognostic factors. The sample consisted of 132 consecutive, surgically resected patient tissues of non-small cell lung cancers, and the following parameters were examined: HER-2/neu and estrogen receptor expression, as well as the related clinical and pathological features: tumor, nodes, and metastases stage, level of tumor necrosis, histological and nuclear grade, lymphocytic infiltrate, and number of mitoses. HER-2/neu was positive in 28.8\% of tumor samples, and estrogen receptor expression was positive in 29.5\% of tumors, but neither was significantly associated with the outcome of non-small cell lung cancers. There was a significant association between HER-2/neu and nuclear grade (P=0.01). In addition, the association between estrogen receptor expression and histological type of tumor (P=0.04) and mitotic rate (P=0.008) was found. Kaplan-Meier analysis showed a significant association of patients' overall survival with the tumor node metastasis stage (P<0.001) and the degree of tumor necrosis (P=0.02). Cox proportional hazard regression analysis showed that male gender (P=0.01), histological type (P=0.03), high degree of necrosis (P=0.006), and higher histological grade (P=0.037) were associated with the patients' survival. Our findings indicate that the expression of HER-2/neu and estrogen receptor is less reliable than traditional histological parameters in predicting the survival of patients with non-small cell lung cancers.

\noindent
\textbf{Case III: A relevant paper which does not rank among top 10}

PubMed ID: 11153605

Title: Ploidy, expression of erbB1, erbB2, P53 and amplification of erbB1, erbB2 and erbB3 in non-small cell lung cancer.

Abstract: The aim of this study was to assess the prognostic value of deoxyribonucleic acid analysis, expression oferbB1, erbB2 and P53, and amplification levels of erbB1, erbB2 and erbB3 in non-small cell lung cancer (NSCLC). Consecutive patients with NSCLC who underwent treatment with curative intention (118) were included. In 108 cases, the cell cycle was analysed using flow cytometry and double-staining with propidium iodide and anticytokeratin. In another 108 cases, expression of erbB1, erbB2 and P53 was assessed immunhistochemically. Amplification of the erbB family was determined in the tumours of 53 patients using double-differential polymerase chain reaction. Of the tumours, 81\% were aneuploid and 14\% showed positive staining for erbB1, 18\% for erbB2 and 41\% for P53. There were normal mean gene copy numbers in 86\% for erbB1, 94\% for erbB2 and in 96\% for erbB3. No significant correlations were noted between erbB1, erbB2 and P53 expression, ploidy status and tumour stage. In a Cox regression model, only tumour stage was shown to be prognostically significant. It seems that ploidy and expression status of erbB1, erbB2 and P53 are not prognostic parameters in non-small cell lung cancer. Amplification of the erbB family does not seem to be a frequent event in non-small cell lung cancer.

\newpage

\bibliographystyle{unsrt}
\bibliography{references.bib}

\end{document}